\documentclass[superscriptaddress,aps,prx,reprint,twocolumn,amsmath,amssymb]{revtex4-2}
\usepackage{graphicx}
\usepackage{hyperref}
\usepackage{amssymb}
\usepackage{slashed}
\usepackage{dcolumn}
\usepackage{amsmath}
\usepackage{bm}
\usepackage{colordvi}
\usepackage{algorithm}
\usepackage{algpseudocode}
\usepackage{multirow}
\usepackage{titlesec}
\usepackage{newtxtext,newtxmath}
\usepackage{dsfont}
\usepackage{xcolor}
\usepackage{mathbbol}
\usepackage{appendix}

\allowdisplaybreaks

\usepackage{mathrsfs}
\makeatletter

\newcommand{\Rmnum}[1]{\expandafter\@slowromancap\romannumeral #1@}
\makeatother

\begin{document}   

\title{Ferroelectric Order and Enhanced Interfacial Superconductivity in Lightly-Doped Quantum Paraelectric KTa$_{1-x}$Nb$_x$O$_3$}

\author{F. Yang}
\email{fzy5099@psu.edu}

\affiliation{Department of Materials Science and Engineering and Materials Research Institute, The Pennsylvania State University, University Park, PA 16802, USA}

\author{L. Q. Chen}
\email{lqc3@psu.edu}

\affiliation{Department of Materials Science and Engineering and Materials Research Institute, The Pennsylvania State University, University Park, PA 16802, USA}

\date{\today}

\begin{abstract}
  Ferroelectric quantum criticality in perovskite oxides offers a fertile ground for emergent collective phenomena. Here we develop a first-principles-inspired  quantum-statistics-based  theoretical analysis of the ferroelectric order and  interfacial superconductivity in lightly-doped quantum paraelectric, niobium (Nb)-doped KTaO$_3$. We demonstrate that local distortions induced by the doped Nb atoms beyond its quantum critical composition induce a long-range ferroelectric order. The predicted dielectric properties quantitatively agree with the experimental measurements over the entire temperature range from the symmetry-broken ferroelectric phase across the phase transition to the paraelectric region. As  the same soft phonon mode that governs dielectric behavior provides the essential  pairing channel for interfacial superconductivity of KTaO$_3$, we predict a pronounced enhancement of this superconductivity on   (111) surface when the system is tuned to its quantum-critical composition via Nb doping, providing a concrete avenue for experimental verification. This finding establishes ferroelectric quantum criticality as a unique design principle for engineering enhanced superconductivity and discovering emergent quantum phases in polar oxide heterostructures,  explicitly suggesting that  similar materials-tuning strategies (e.g., epitaxial strain) could be exploited to enhance superconductivity in quantum paraelectric systems.
\end{abstract}

\maketitle  

{\sl Introduction.---}The long-range ordering in quantum materials has long been a central theme in materials science and condensed matter physics, with far-reaching implications across the correlated-electron and functional oxide systems.  Long-range order develops through extended correlations mediated by massless/gapless collective excitations, which arise upon condensation in many-body systems and  mediate the exchange interactions between distant parts of the system or between constituent particles~\cite{nambu2009nobel, auerbach2012interacting}. These excitations are typically associated with spontaneously broken symmetries in systems like superconductors, magnets, or superfluids, and the existence of such gapless modes is a hallmark of systems near a critical point, where the system is at a phase transition between different states of matter. In general, local lattice irregularities or compositional fluctuations that break translational symmetry tend to destroy long-range order. This occurs because local perturbations scatter the excitations,  disrupting their propagation (i.e., reducing their lifetime) and reducing the correlation length~\cite{abrikosov2012methods,mahan2013many,auerbach2012interacting}. When sufficiently strong, local perturbations can even pin the collective modes~\cite{fukuyama1978dynamics,gruner1985charge,gruner1988dynamics}, leading to the opening of an excitation gap, the localization of excitations, and the suppression of long-range correlations.  

In certain quantum materials, however, local perturbations can promote cooperative phenomena, 
stabilizing or even inducing long-range order and giving
rise to emergent phase.  A paradigmatic example arises in displacive quantum paraelectrics such as SrTiO$_3$ and KTaO$_3$~\cite{muller1979srti, rowley2014ferroelectric, cowley1996phase, fujishita2016quantum, verdi2023quantum, yang2024microscopic,singh1996stability,wu2022large,li2019terahertz,cheng2023terahertz,74d5-4hsw}. These systems possess intrinsic lattice-dynamical instabilities that would have led to paraelectric-to-ferroelectric transitions at low temperatures if quantum effects were absent~\cite{muller1979srti,verdi2023quantum,yang2024microscopic,wu2022large}. However, strong zero-point oscillation of the lattice vibrations renormalizes the unstable soft-mode dynamics and suppresses the ferroelectric transition~\cite{verdi2023quantum,yang2024microscopic,wu2022large}, yielding an incipient ferroelectric ground state. In such a delicately balanced  state, even subtle compositional tuning or site substitution~\cite{andrews1985x,toulouse1992precursor,aktas2014polar,rischau2017ferroelectric,takesada2006perfect,rowley2014ferroelectric}  may profoundly alter the ground state of the entire system. For example, as little as 0.8\% Nb substitution in KTaO$_3$ is sufficient to drive the system from a quantum paraelectric to a ferroelectric phase in the zero-temperature limit~\cite{10.1103/physrevlett.39.1158}.

\begin{figure}
    \includegraphics[width=8.7cm]{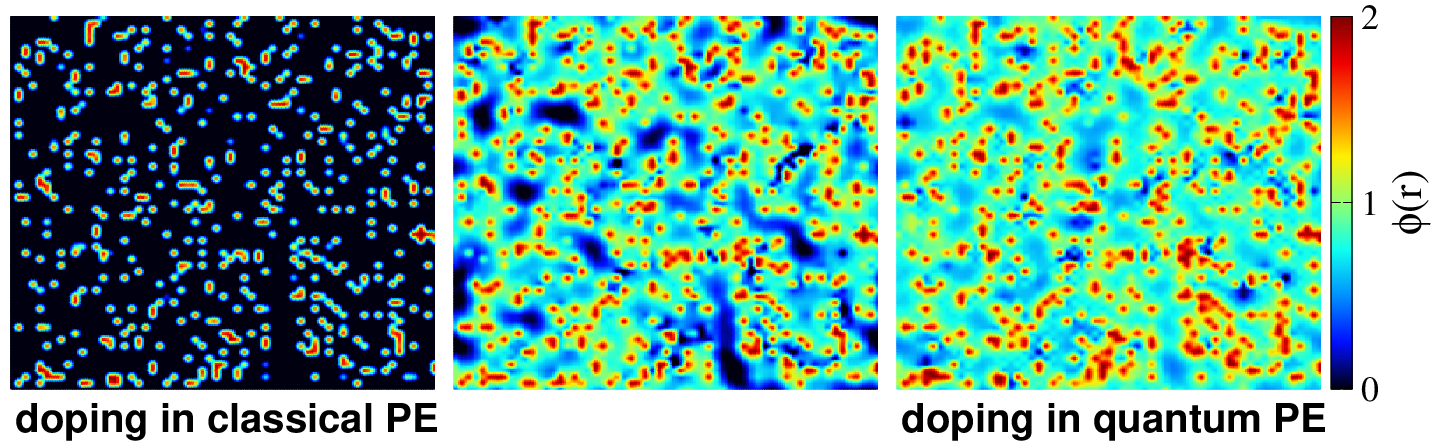}
    \caption{Schematic illustration of the ground state in the presence of random  local
lattice distortions, showing the spatial distribution of the order parameter (lattice distortion $\phi$). Left and right panels correspond to the doping cases in a classical paraelectric (far away from ferroelectric instability) and a quantum paraelectric (on the verge of ferroelectric instability), respectively, while the middle panel denotes an intermediate case. When impurities seed local polar regions, dilute dipolar distortions cannot drive classical paraelectric into globally ferroelectric state and remain isolated and uncorrelated. In quantum paraelectric, the nearly gapless soft phonon mode efficiently mediates long-range correlations, triggering a collective condensation of the host lattice into globally ferroelectric phase.}
    \label{figyc1}
\end{figure}

The microscopic origin of this transition has recently attracted much  attention~\cite{Klossner01071994,Maglione01121992,PhysRevB.53.5240,doi:10.1073/pnas.2419159122,zhao2025observation,Maglione_1987}, as Nb substitution in KTaO$_3$ does not introduce extrinsic itinerant charge carriers yet profoundly modifies the dynamics of low-lying  excitations. Moreover, the low doping level of 0.8\% Nb is not sufficient to produce periodic or compositionally ordered structure, i.e., the Nb atoms act as randomly distributed, dilute point defects embedded in the KTaO$_3$ host lattice. First-principles calculations at the atomic scale~\cite{doi:10.1073/pnas.2419159122} have revealed that Nb substitution induces local lattice distortions and off-center displacements of Nb ions within the unit cell, generating {\sl local} dipolar moments. These dipoles act as nucleation centers for polarization coupled strongly to the transverse-optical soft phonon modes of the host lattice. The  quantum paraelectric on the verge of ferroelectric instability is characterized by a nearly gapless soft phonon mode~\cite{muller1979srti, rowley2014ferroelectric, cowley1996phase, fujishita2016quantum, verdi2023quantum,wu2022large,li2019terahertz,cheng2023terahertz,74d5-4hsw,doi:10.1073/pnas.2419159122}, and hence, this mode, in principle, can effectively mediate the long-range ferroelectric correlations across distant dipoles, in contrast to the classical paraelectrics with a fully gapped polar spectrum. Thus, in one view, as schematically illustrated in Fig.~\ref{figyc1} and also realized by  experiment~\cite{PhysRevB.53.5240}, once Nb impurities seed local polar puddles, even dilute defect  dipolar distortions can trigger a collective condensation of lattice distortion mediated by soft phonons, driving the quantum-paraelectric host into a globally polarized ferroelectric phase and establishing long-range order.

Capturing such a collective behavior, however, lies beyond the reach of the first-principles  atomistic simulations, which can resolve local distortions but fail to describe the emergent mesoscale-to-macroscopic collapse of the host lattice into a ferroelectric state.  Here we perform a first-principles-inspired analysis of Nb-doped KTaO$_3$ by combining a bosonic-condensation framework with a quantum-statistical self-consistent renormalization approach~\cite{abrikosov2012methods,mahan2013many}. This treatment 
 couples Nb-induced local
lattice distortions with the long-range bosonic condensation of soft phonons and captures the resulting ferroelectric order in quantum paraelectrics while simultaneously accounting for the ferroelectric order and its thermal fluctuations across the full temperature regime. 
  In addition to explaining the induced ferroelectric transition, the predicted dielectric behaviors at different Nb doping levels quantitatively agree with existing experimental measurements  over a wide temperature range from the zero-temperature ferroelectric ground state all the way to the paraelectric state. Furthermore, the present framework also provides a natural basis for understanding other phenomena governed by soft-phonon condensation, particularly the enhancement of the superconducting transition temperature for interfacial  superconductivity on the KTaO$_3$(111) surface by tuning system to its quantum-critical composition via Nb doping. 

{\sl Bosonic-condensation.---}We focus exclusively on the soft transverse optical (TO) phonon mode, as it governs the displacive ferroelectric instability of KTaO$_3$~\cite{cheng2023terahertz}. The longitudinal optical phonons are neglected because they are strongly stiffened by the long-range Coulomb interaction~\cite{palova2009quantum,PhysRevLett.131.046801,PhysRevLett.72.3618,ashcroft1976solid,yang2024microscopic} and thus remain inactive. The coupling of acoustic phonons to the soft TO mode can be treated perturbatively~\cite{rowley2014ferroelectric,palova2009quantum,yang2024microscopic} and does not qualitatively influence the condensation behavior caused by soft TO phonons. These approximations are well justified in describing the low-lying lattice dynamics of perovskite quantum paraelectric KTaO$_3$. The field operator associated with the soft TO phonon is denoted by $\phi_{\rm sp}({\bf r})$. In the lattice-dynamical expansion up to quartic order, the effective Hamiltonian of soft TO phonon in KTa$_{1-x}$Nb$_x$O$_3$ reads
\begin{align}
  &H=\frac{1}{2}\int{d{\bf r}}\phi^*_{\rm sp}({\bf r})\Delta^2_{\rm sp}(x)\phi_{\rm sp}({\bf r})+\frac{1}{4}\int\!\!{d{\bf r}}u_b|\phi_{\rm sp}({\bf r})|^4\nonumber\\
  &\mbox{}+\frac{1}{2}\int\!\!{d{\bf r}}{d{\bf r'}}\phi^*_{\rm sp}({\bf r})Q({\bf r\!-\!r'})\phi_{\rm sp}({\bf r}')\!-\!\int\!\!d{\bf r}{\bf h}_{\rm eff}\cdot{\bf e}_{\rm sp}({\bf r}),\label{PHAM}
\end{align}
where  
\begin{equation}
  \Delta^2_{\rm sp}(x)=\Delta^2_{\rm sp}(0)-\Sigma_{\rm imp}({\bf r}).
\end{equation}
Here,  $u_b$ is the coupling strength of the local quartic self-interaction;  $Q({\bf r}-{\bf r'})$ represents the nonlocal elastic restoring interaction responsible for the phonon dispersion with the Fourier transform $Q({\bf q})=v^2q^2$ ($v$ being the phonon mode velocity); $\Delta_{\rm sp}(0)$ is the gap of the soft phonon mode; $\Sigma_{\rm imp}({\bf r})=\sum_i\delta U\delta({\bf r}-{\bf r}_i)$ describes the local self-energy arising from dopant-induced lattice distortions~\cite{doi:10.1073/pnas.2419159122} through electrostrictive coupling~\cite{lines2001principles}, where $\delta U$ characterizes the local potential strength associated with a dopant at position ${\bf r}_i$;  ${\bf e}_{\rm sp}({\bf r})$ denotes the unit vibration-direction vector of the local soft-mode displacement, and ${\bf h}_{\rm eff}$ represents a symmetry-breaking field that couples with the soft-mode displacement, arising from oriented defect dipoles or a built-in electric field induced by doping, which breaks the cubic symmetry and favors polarization alignment along specific crystallographic directions.

Under the assumption of random disorder distribution and dilute dopant levels~\cite{mahan2013many,abrikosov2012methods}, averaging over impurity positions yields an effective static correction to the phonon self-energy, $\Sigma_{\rm imp}({\bf r}) = -x\delta\bar U$, where $x$ denotes the dopant composition. The soft-phonon gap then takes the form
\begin{equation}\label{omega0}
\Delta_{\rm sp}^2(x)=\Delta_{\rm sp}^2(0)-x{\delta \bar U}.
\end{equation}
The soft phonon mode acquires an imaginary frequency when $\Delta_{\rm sp}^2(x)$ becomes negative, signaling the onset of a bosonic condensation of the soft-phonon field at the zone center ($q=0$). This collective condensation of the soft-phonon mode gives rise to a nonzero macroscopic displacement field ${\hat \xi}=\langle|{\phi^{*}_{\rm sp}(q=0)}|\rangle$~\cite{lee1974conductivity}  that couples directly to the polarization field ${\bf \hat P}=({z^*{\bf u}}/{\Omega_{\rm cell}}){\hat \xi}$ in a displacive ferroelectric crystal~\cite{cochran1981soft,cochran1961crystal,cowley1996phase,cowley1965theory,cochran1969dynamical,yelon1971neutron}, where $\mathbf{u}$ is the ionic vibration vector, $z^*$ is the effective charge, and $\Omega_{\rm cell}$ denotes the unit cell volume. This marks the emergence of a ferroelectric state in the ground state.

Focusing on the long-wavelength regime where the collective lattice distortion develops into a spatially coherent polarization field, the resulting mean-field Hamiltonian 
 of the polarization field to describe the emergent long-range (global) ferroelectric order and its spatial variation takes the form
\begin{equation}\label{Ham}
  \mathcal{H}_0=\!\!\int\!\!{d{\bf r}}\Big[\frac{g}{2}({\bf \nabla}{\hat P})^2+\frac{a_0(x)}{2}{\hat P}^2+\frac{b_0}{4}{\hat P}^4-\kappa\cos\theta\Big],
\end{equation}
where 
\begin{equation}
a_0(x) = a_0(0) - x\,\delta a \propto\Delta^2_{\rm sp}(0)-x\delta {\bar U} 
\end{equation}
represents the quadratic coefficient, $b_0\propto{u_b}$ is the anharmonic coefficient,  $g\propto{v^2}$ characterizes the gradient stiffness~\cite{lines2001principles}, and $\kappa\propto{h_{\rm eff}}$ quantifies the strength of the defect–polarization orientation coupling and determines the stability of the preferred polarization direction, while $\theta$ denotes the angle between the polarization and the [100] crystallographic axis.  

It is important to note that in many-body physics~\cite{abrikosov2012methods,mahan2013many,peskin2018introduction}, the zero-point renormalization of lattice vibrations does not alter the analytical structure of Hamiltonians and merely renormalizes its model parameters. Here we consider that the Hamiltonian (\ref{Ham}) has already incorporated this renormalization effect~\cite{yang2024microscopic}. Consequently, all parameters appearing here are the renormalized zero-temperature values rather than the bare ones. Under this condition, $a_0(x=0)$ for a quantum paraelectric remains positive but nearly vanishing~\cite{yang2024microscopic,rowley2014ferroelectric,palova2009quantum}, placing the system close to a ferroelectric instability.

\begin{figure}
    \includegraphics[width=8.5cm]{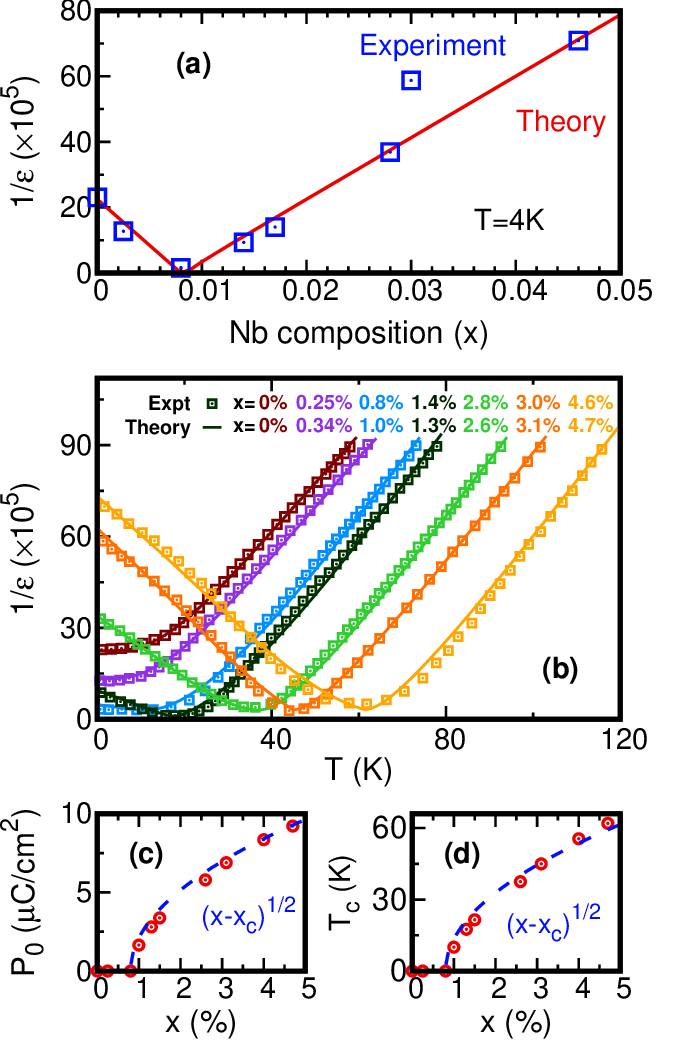}
    \caption{Dielectric behavior of Nb-doped KTaO$_3$.
(a) Inverse dielectric function $1/\varepsilon$ at the low-temperature limit for various Nb compositions.
      (b) Temperature dependence of $1/\varepsilon$ at different Nb doping levels. The solid curves in (a) and (b) represent the theoretical results, while the experimental data (squares) are taken from Ref.~\cite{10.1103/physrevlett.39.1158}. In panel (b), because the experimentally reported Nb compositions may contain calibration uncertainties, doping compositions in our theoretical calculations are adjusted slightly by aligning $1/\varepsilon(T=0,x)$ to match the experimental data. (c) Zero-point polarization and (d) transition temperature as functions of Nb concentration. The circles correspond to our calculated results, and the dashed lines represent the empirical relations proposed experimentally~\cite{10.1103/physrevlett.39.1158}. We set $\phi = 0.304\pi$, corresponding to polarization along the [100] direction and the measurement field applied along [111] direction. }    
\label{figyc2}
\end{figure}

At the low-temperature limit, the dielectric behavior is primarily controlled by the doping dependence of the quadratic coefficient $a_0(x)$. The zero-point dielectric constant reads
\begin{equation}
\varepsilon(T=0,x)=1+\frac{1/\varepsilon_0}{a_0(x)+b_0P_0^2(x)(1+2\cos^2\phi)}, 
\end{equation}
where the zero-point polarization amplitude satisfies $P_0^2(x)=0$ for $a_0(x)>0$ (paraelectric phase) and $P_0^2(x)=-a_0(x)/b_0$ for $a_0(x)<0$ (ferroelectric phase), indicating that a sign reversal of $a_0(x)$ across the critical composition $x_c=\delta{a}/a_{0}(0)$ signals the transition from paraelectric to ferroelectric ground state. Here, $\phi$ denotes the angle between the polarization direction and the probing electric field. Consequently, one has 
\begin{equation}
  \varepsilon(T\sim0,x)=\begin{cases}
  1+\frac{1/\varepsilon_0}{a_0(x)},~~x<x_c, \\
  1-\frac{1/\varepsilon_0}{2a_0(x)\cos^2\phi},~~x>x_c.
  \end{cases}
\end{equation}
Using the experimentally determined parameter $a_0(0)$ of pristine KTaO$_3$~\cite{yang2024microscopic,rowley2014ferroelectric} together with the quantum critical composition $x_c=0.008$~\cite{10.1103/physrevlett.39.1158}, the calculated zero-temperature dielectric constant is shown in Fig.~\ref{figyc2}(a). Without fitting parameters, the theoretical results reproduce the experimental data at $4~$K~\cite{10.1103/physrevlett.39.1158} with excellent quantitative agreement.

{\sl Formulation of Dielectric properties.---}The Hamiltonian (\ref{Ham}) characterizes the ground-state (zero-point) properties of the system and provides the foundation for describing the polarization configuration.  Theories for different classes of ordered systems follow distinct microscopic paths from their ground-state Hamiltonian to their effective macroscopic descriptions at finite temperatures, depending sensitively on the nature of the underlying order. In superconductors or ferromagnets, ground-state theories such as the BCS or Heisenberg models yield quasiparticle description (Bogoliubov quasiparticles or magnons) whose thermal populations self-consistently reproduce the finite-temperature phase behavior and the critical phenomena, and near the critical temperature $T_c$, can successfully reduce to the polynomial Ginzburg–Landau forms.  For displacive ferroelectrics: its first-principle  ground-state mean-field Hamiltonian is inherently polynomial, reflecting a built-in structural instability of the lattice that  encodes the dynamics of the unstable phonon mode. Definitely, it is fundamentally different from the phenomenological Ginzburg–Landau free energy, which is constructed as a series expansion in terms of order parameters under the assumption that the magnitudes of the order parameters are small near the critical temperatures and requires a known knowledge of critical behaviors and temperature-dependent parameter fitting. By contrast, a quantum statistic treatment can, in principle,   extend this Hamiltonian to a macroscopic model at finite temperatures, solve the phase-transition criticalities (e.g., $T_c$)  {\sl without} the needs of phenomenological fitting, and near the critical temperature $T_c$, can recover the polynomial Ginzburg–Landau forms.
 
We employ a thermodynamic treatment within the quantum statistic mechanics~\cite{abrikosov2012methods,mahan2013many,yang2024microscopic}. We decompose the polarization field as $\hat{\bf P} = \mathbf{P} + \delta\mathbf{P}_{\rm th}(T)$, where $\mathbf{P}$ is the average polarization, and $\delta\mathbf{P}_{\rm th}(T)$ denotes the thermal fluctuations. We introduce a stochastic thermal field ${\bf E}_{\rm th}(t,{\bf R})$ that obeys the fluctuation-dissipation theorem~\cite{tang2022excitations,landau1981statistical}: $\langle{\bf E}_{\rm th}\rangle=0$ and 
\begin{equation}
\langle{\bf E}_{\rm th}(\omega,{\bf q}){\bf E}_{\rm th}(\omega',{\bf q}')\rangle=\frac{\gamma\hbar\omega(2\pi)^4\delta({\bf q}\!-\!{\bf q'})\delta(\omega\!-\!\omega')}{\tanh(\beta\hbar\omega/2)},  
\end{equation}
with $\gamma=0^+$ and $\beta=1/(k_BT)$. The resulting dynamics of polarization fluctuations becomes (See Supplement Materials)
\begin{equation}
\big[m_p\partial_t^2+\gamma\partial_t-g\nabla^2+m_p\Delta^2\big(P^2,\langle\delta{P}^2\rangle\big)\big]\delta{\bf P}={\bf E}_{\rm th}(t,{\bf R}),
\end{equation}
with the excitation gap determined by
\begin{align}
&m_p\Delta^2\!=\!a_0(x)\!+\!b_0(2/3\!+\!1){P}^2\!+\!b_0(2/3\!+\!1)\langle\delta{P}^2\rangle.\label{gap}
\end{align}
Here, $m_p$ is the polarization inertia. As a result, the thermally averaged $\langle\delta{\bf P}_{\rm th}(T)\rangle$  vanishes whereas the mean-square fluctuation $\langle{\delta{P^2_{\rm th}(T)}}\rangle$ remains finite in the thermal ensemble,
\begin{equation}
  \langle{\delta{P}^2_{\rm th}}\rangle=\!\!\int\!\!\frac{{\hbar}d{\bf q}}{m_p(2\pi)^3}\frac{n_B[\hbar\omega_{\bf q}(T,x)]}{\omega_{\bf q}(T,x)}.\label{tf}
\end{equation}
These fluctuations arise from the bosonic thermal excitation $n_B(\hbar \omega_{\bf q})$ of a collective polar mode, whose energy dispersion:  
\begin{equation}
\hbar \omega_{\bf q}(T, x) = \hbar \sqrt{ \Delta^2\big(P^2,\langle\delta{P}^2\rangle\big)+{g q^2}/{m_p} }.
\end{equation}
Substituting $\langle\delta{\bf P}_{\rm th}(T)\rangle=0$ and the finite mean-square fluctuation $\langle{\delta{P^2_{\rm th}(T)}}\rangle$ to the Hamiltonian (\ref{Ham}) and extracting the part related to $P^2$ yield the effective free energy for the long-range (global) ferroelectric ordering
\begin{equation}\label{freeenergy}
\mathcal{F}=\int{d{\bf x}}\Big[\frac{\alpha{(T,x)}}{2}P^2+\frac{b_0}{4}P^4\Big]+\mathcal{F}_{an}(\theta),
\end{equation}
where the renormalized quadratic coefficient takes the form
\begin{equation}
  \alpha(T,x)=a_0(x)\!+\!b_0(2/3+1)\langle\delta{P}^2_{\rm th}(T)\rangle.\label{alpha}
\end{equation}
Here, $\mathcal{F}_{\rm an}(\theta)$ accounts for the orientational anisotropy of polarization induced by $\kappa(1-\cos\theta)$; at low temperatures, it favors the polarization along the [001] direction, while at higher temperatures, thermal rotation of the polarization reduces the effective anisotropy. Minimization of this free energy yields a ferroelectric phase ($P^2=-\alpha(T,x)/b_0$) when $\alpha(T,x) < 0$ and a paraelectric phase ($P^2 = 0$) when $\alpha(T,x) > 0$.

By solving Eq.~(\ref{alpha}) self-consistently, we determine $\alpha(T,x)$ and thus the dielectric behavior at finite temperature. The relative permittivity as a function of Nb concentration reads
\begin{equation}
\varepsilon(T,x)=1+\frac{1/\varepsilon_0}{\alpha(T,x)+b_0P^2(T,x)G(T)}, 
\end{equation}
where the geometrical–orientational factor $G(T)$ accounts for the thermal distribution of polarization orientations,
\begin{eqnarray}
  G\!\!&=&\!\!1\!+\!2\langle\cos^2(\theta-\phi)\rangle=1\!+\!2\frac{\int\sin\theta{d}\theta\cos^2(\theta\!-\!\phi)e^{\beta\kappa\cos\theta}}{\int\sin\theta{d}\theta{e^{\beta\kappa\cos\theta}}}\nonumber\\
  &=&\!2-W(T)+(3W(T)-1)\cos^2\phi,  
\end{eqnarray}
with $W(T)=\langle\cos^2\theta\rangle=1-\frac{2}{\beta\kappa}[\coth(\beta\kappa)-\frac{1}{\beta\kappa}]$.

Consequently, this self-consistent formulation of the
vectorial fluctuations and global polarization starting from the ground state enables one to use only the ground-state parameters to predict the dielectric/ferroelectric
properties at finite temperatures at different Nb doping. The coefficients $b_0$ and $\kappa$ are also expected to vary with the level of doping, with $b_0$ containing its effects of strongly local and anharmonic lattice distortions and and $\kappa$ reflecting the interactions among off-centered dipoles. Reliable calculation of their evolution with composition is challenging as it lies beyond the scope of conventional first-principles and self-energy-renormalization (Born-approximation-level) theories.  In this work, for each doping level, we determined $b_0$ and $\kappa$ by fitting to experimental data~\cite{10.1103/physrevlett.39.1158} in the low-temperature regime ($T<20$~K), where thermal effects of collective excitations are negligible. Based on the obtained values for these parameters of the ground state, it is able to quantitatively predict the temperature-dependent dielectric behavior of the system at higher temperatures, including the phase-transition criticalities and beyond.

The calculated dielectric properties at different Nb compositions as a function of temperature from the zero-temperature condensed ground state to the paraelectric region are shown in Fig.~\ref{figyc2}(b). The theoretical curves quantitatively reproduce the experimentally measured dielectric properties~\cite{10.1103/physrevlett.39.1158} over the entire temperature range, showing excellent agreement with the experimental data. The computed doping dependencies of both the zero-point polarization $P_0(x)$ and the ferroelectric transition temperature $T_c(x)$ are shown in Figs.~\ref{figyc2}(c) and (d), respectively. They closely follow the empirical power-law relations $P_0(x)\propto\sqrt{x-x_c}$ and $T_c(x)\propto\sqrt{x-x_c}$ previously proposed based on experiments~\cite{10.1103/physrevlett.39.1158}. 
These agreements directly confirm our theoretical hypothesis that the Nb-induced ferroelectric transition in KTaO$_3$ originates from the collective condensation of the soft-phonon field, where local dipolar distortions seeded by dilute Nb dopants cooperate to transform quantum-paraelectric host into long-range ferroelectric state.

{\sl Application to Interfacial Superconductivity.---}We now apply the above framework to propose an experimentally testable consequence  governed by the soft-phonon condensation. Recent experimental and theoretical studies~\cite{10.1038/s41467-023-36309-2} have shown that low-lying soft phonons in KTaO$_3$ play a decisive role in the mediating superconducting pairing of 2D electron gap formed at polar oxide interfaces (e.g., YAlO$_3$/KTaO$_3$~\cite{Zhang2023}, LaAlO$_3$/KTaO$_3$~\cite{PhysRevX.15.021018,10.1038/s41467-024-51969-4,10.1126/science.abb3848,10.1103/physrevlett.126.026802,10.1126/science.aba5511,PhysRevX.15.011037,PhysRevX.15.011006,dong2025strongly,zhang2025universally,1848-9r5d}, AlO$_x$/KTaO$_3$~\cite{10.1038/s41467-022-32242-y}, EuO/KTaO$_3$~\cite{10.1126/science.aba5511}, LaTiO$_3$/KTaO$_3$~\cite{10.1063/5.0151227} and CaZrO$_3$/KTaO$_3$~\cite{Zhang2025,chen2025two}), i.e.,  the same soft phonon mode governing the dielectric behavior also serves as efficient mediator for interfacial superconductivity in quantum paraelectrics.  Specifically, within the BCS framework~\cite{bardeen1957microscopic,schrieffer1964theory}, the onset temperature of superconductivity is
\begin{equation}\label{SP}
  T^{\rm SC}_{\rm os}=1.134\Omega_{c}\exp(-1/\lambda),
\end{equation}
where $\lambda = -D\langle g_{\mathbf{kk'}}\rangle_F$ is the dimensionless coupling constant with $D$ being the electron density of states, $\Omega_c$  the cutoff energy of the attractive interaction, $\langle\dots\rangle_F$  the Fermi-surface average. The effective pairing potential is given by~\cite{bardeen1957microscopic,schrieffer1964theory,abrikosov2012methods,mahan2013many} 
\begin{equation}\label{SCpp}
  g_{\bf kk'}=\frac{{\bar \omega}_{\bf k-k'}|D^{\rm ph-e}_{\bf k-k'}|^2}{(\xi_{{\bf k},n}-\xi_{{\bf k'},n'})^2-{\bar \omega}^{2}_{\bf k-k'}}.
\end{equation}  
Here, ${\bar \omega}_{\mathbf{k}-\mathbf{k'}}$ is the characteristic phonon frequency mediating the interaction, $\xi_{{\bf k},n}=\varepsilon_{{\bf k},n}-\mu$ and $\varepsilon_{{\bf k},n}$ denotes the energy of band $n$. The electron–phonon coupling matrix element is $D^{\rm ph-e}_{\bf k-k'}=\sum_{\mu}{{\bf e}_{\mu}\cdot({\bf k-k'})}U^{n,n'}_{{\bf k},{\bf k'}}/{\sqrt{2M_{\mu}{\bar \omega}_{\bf k-k'}}}$, with ${\bf e}_{\mu}$ being the vibrational eigenvector of the $\mu$-th ion of mass $M_{\mu}$ in the unit cell, and $U^{n,n'}_{{\bf k},{\bf k'}}$ the lattice potential matrix element.

\begin{figure}
    \includegraphics[width=8.7cm]{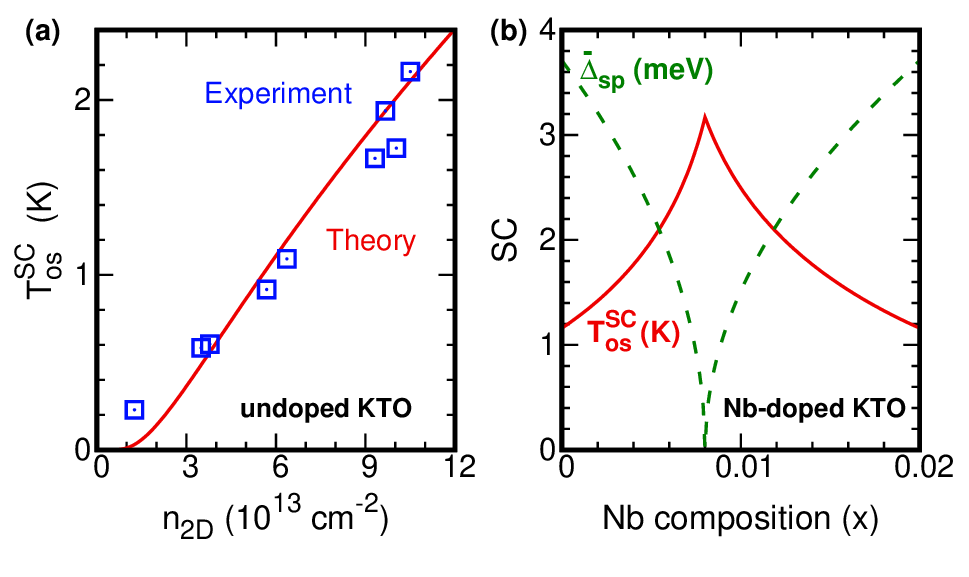}
    \caption{(a) Carrier density dependence of the superconducting onset temperature $T_{\rm os}^{\rm SC}$ of the 2DEG formed at undoped KTaO$_3$ (111) interface. Experimental data are taken from Ref.~\cite{10.1038/s41467-023-36309-2} for EuO/KTaO$_3$ (111) interface. {\bf b} Theoretical prediction of  $T_{\rm os}^{\rm SC}$ (solid curve) and renormalized soft-phonon gap (dashed curve) as a function of Nb doping for interfacial superconductivity at the EuO/KTa$_{1-x}$Nb$_x$O$_3$(111) interface, showing a pronounced maximum and minimum   near the quantum critical composition ($x=0.8\%$), respectively.}    
\label{figyc3}
\end{figure}

In conventional superconducting metals, electron pairing is primarily mediated by acoustic phonons through lattice vibrations~\cite{bardeen1957microscopic,schrieffer1964theory}. However, in quantum paraelectrics, this contribution becomes weak because of high sound velocities and reduced electron–acoustic-phonon coupling, while the low-lying soft optical phonon mode opens an additional, dominant pairing channel~\cite{10.1038/s41467-023-36309-2}. Accordingly, ${\bar \omega}_{\mathbf{k}-\mathbf{k'}}$ in Eq.~(\ref{SCpp}) represents the soft-phonon dispersion, which takes the form ${\bar \omega}_{\bf q}=\sqrt{{\bar \Delta}^2_{\rm sp}+v^2q^2}$. As shown in Ref.~\cite{10.1038/s41467-023-36309-2}, this pairing mechanism naturally explains the orientation selectivity~\cite{10.1038/s41467-023-36309-2,10.1126/science.aba5511} and the anomalous linear scaling of $T_{\rm os}^{\rm SC}$ with carrier density~\cite{10.1038/s41467-023-36309-2} observed in KTaO$_3$-based 2D electron gases. Specifically, owing to symmetry lowering, different crystallographic orientations project distinct components of the soft-phonon vibrations onto the interfacial electron gas. The geometric overlap between the phonon polarization vector ${\bf e}_{\mu}$ and the in-plane momentum transfer ${\bf k-k'}$ determines the pairing potential $g_{\bf kk'}$, being maximal for the (111) interface, reduced for (110), and minimal for (001). This explains why experimentally observed superconductivity is robust on the (111) interface, weakens on (110), and is absent on (001)~\cite{10.1038/s41467-023-36309-2,10.1126/science.aba5511}. Moreover, evaluating Eq.~(\ref{SP}) and Eq.~(\ref{SCpp}) yields a simplified form
\begin{equation}\label{SCTC}
  T^{\rm SC}_{\rm os}=1.134\Omega_{c}\exp\big[-1/(c_0\langle|{\bf k-k'}|^2/{\bar \omega}^{2}_{\bf k-k'}\rangle_F)\big],  
\end{equation}
with $c_0$ being a constant. Assuming approximate degeneracy of the three $t_{2g}$ orbitals at the KTaO$_3$(111) interface, the Fermi wavevector is $k_F=\sqrt{2\pi n_{\rm 2D}/3}$, with $n_{\rm 2D}$ the interfacial carrier density. Moreover, in quantum paraelectric KTaO$_3$, ${\bar \Delta}_{sp}=3.79$~meV~\cite{cheng2023terahertz,74d5-4hsw}. We take the cutoff $\Omega_c={\bar \omega}_{q=k_F}$. Then, as shown in Fig.~\ref{figyc3}(a), the pairing mediated by the low-lying soft phonon mode naturally yields a linear increase of $T_{\rm os}^{\rm SC}$ with carrier density $n_{\rm 2D}$, quantitatively  consistent with the existing experimental data on EuO/KTaO$_3$(111) interfaces~\cite{10.1038/s41467-023-36309-2}.

It is noted from Eq.~(\ref{SCTC}) that the interfacial superconductivity is enhanced as the soft-phonon mode softens. In the framework of the present study, Nb doping in KTaO$_3$ drives the soft phonon toward condensation. Then, the renormalized soft-phonon gap ${\bar \Delta}_{\rm sp}(x)$ exhibits the same doping dependence as the polar-mode gap [Eq.~(\ref{gap})], and thus in the low-temperature limit one has ${\bar \Delta}^2_{\rm sp}(x) = {\Delta}^2_{\rm sp}(x)$ for $x < x_c$ and ${\bar \Delta}^2_{\rm sp}(x) = -\frac{2}{3}{\Delta}^2_{\rm sp}(x)$ for $x > x_c$, i.e., ${\bar \Delta}_{\rm sp}(x)$ coincides with ${\Delta}_{\rm sp}(x)$ [Eq.~(\ref{omega0})] in the paraelectric phase but changes sign and magnitude in the ferroelectric phase due to the inversion of the soft-mode curvature. Consequently, the soft phonon becomes nearly gapless near the quantum critical Nb composition, ${\bar \Delta}_{\rm sp}(x\rightarrow0.8\%)\to0$, thereby maximizing the superconducting pairing interaction. At a fixed $n_{\rm 2D}=6\times10^{13}/\text{cm}^2$, substituting the calculated ${\bar \Delta}_{\rm sp}(x)$ into Eq.~(\ref{SP}) yields, as shown in Fig.~\ref{figyc3}(b), a pronounced and sharply peaked $T_{\rm os}^{\rm SC}(x)$ that reaches approximately 3~K precisely at the ferroelectric quantum critical point ($x=0.8\%$). This peak value is nearly twice $T_{\rm os}^{\rm SC}(x=0)$ of the undoped system. 
These results demonstrate that the ferroelectric quantum criticality provides an efficient route to enhance superconductivity at oxide interfaces, and highlights the tunability of the soft-phonon-mediated pairing, suggesting that similar strategies such as chemical substitution~\cite{andrews1985x,toulouse1992precursor,aktas2014polar,rischau2017ferroelectric,takesada2006perfect,rowley2014ferroelectric} or epitaxial strain~\cite{xu2020strain,li2025classical} could be exploited to boost  superconductivity in quantum paraelectric systems. They further indicate that ferroelectric order and superconductivity can coexist on the ferroelectric side of the quantum critical point ($x>0.8\%$), establishing an exotic regime of coupled polar and superconducting phases~\cite{zhang2025universally,dong2025strongly}. 

{\sl Summary.---}We formulated a first-principles-inspired  analysis by combining a bosonic-condensation description of the soft phonons with a quantum-statistical self-consistent renormalization approach to elucidate the microscopic origin of Nb-doping-induced ferroelectricity and enhanced interfacial superconductivity in quantum paraelectric KTaO$_3$. The resulting formulation captures the collective condensation of the soft TO phonon mode driven by Nb-induced local lattice distortions, and reproduces the full dielectric/ferroelectric behavior of KTa$_{1-x}$Nb$_x$O$_3$ across wide ranges of temperature and levels of doping. Beyond accounting for the ferroelectricity, it is demonstrated that the same framework naturally extends to interfacial superconductivity of the 2DEG 
formed at the polar oxide interface, where the soft phonon mode serves as an efficient pairing mediator. The model predicts a sharp enhancement of the superconducting onset temperature $T_{\rm os}^{\rm SC}$ near the ferroelectric quantum critical point, directly linking the soft-phonon condensation to the pairing mechanism.  These findings establish a coherent microscopic connection between ferroelectric quantum criticality and superconductivity, showing that the collective collapse of a soft phonon field in a quantum paraelectric host can generate both macroscopic polarization and enhanced electron pairing.

{\it Acknowledgments.---}We acknowledge valuable and insightful discussions with J. Kim, M.Q. Yu, J.F. Yang, J. Levy and C.B. Eom. This work is supported by the US Department of Energy, Office of Science, Basic Energy Sciences, under Award Number DE-SC0020145 as part of Computational Materials Sciences Program.  F.Y. and L.Q.C. also appreciate the generous support from the Donald W. Hamer Foundation through a Hamer Professorship at Penn State.

\begin{widetext}
\begin{appendix}
\renewcommand{\thesection}{Appendix~\Roman{section}} 
\renewcommand{\theequation}{S\arabic{equation}}
\renewcommand{\thetable}{S\Roman{table}}
\renewcommand{\thefigure}{S\Roman{figure}}
 
\section{Displacive ferroelectricity}

\subsection{Derivation of the displacive ferroelectric  model}

Here we present the detailed microscopic derivation of the self-consistent renormalization model for the displacive ferroelectrics~\cite{yang2024microscopic}. Starting from the mean-field Hamiltonian of the polarization field in the main text, which is derived from the underlying lattice dynamics of the unstable soft-phonon mode,  the ground-state Lagrangian for polarization field takes the form:
\begin{equation}\label{SHam}
\mathcal{L}=\frac{m_p}{2}(\partial_t{\hat P})^2-\Big[\frac{g}{2}({\bf \nabla}{\hat P})^2+\frac{a}{2}{\hat P}^2+\frac{b}{4}{\hat P}^4-\kappa\cos\theta\Big].
\end{equation}
For completeness, the model parameters $a$ and $b$ here are the bare parameters. We decompose the polarization field as $\hat{\bf P} = \mathbf{P} + \delta\mathbf{P}$ and assume isotropic fluctuations. The odd-order fluctuation terms vanish under thermal average, and then, the Lagrangian becomes: 
\begin{equation}\label{SLAG}
\mathcal{L}\!=\!\frac{m_p}{2}(\partial_t\delta{P})^2-\big[\frac{g}{2}({\bf \nabla}\delta{P})^2+\frac{a}{2}{\delta}P^2+\frac{b}{4}{\delta}P^4\big]-\big(\frac{a}{2}P^2+\frac{b}{4}P^4\big)-\frac{b}{2}\big(\frac{2}{d_p}+1\big){P}^2\delta{P}^2+\kappa\cos\theta,
\end{equation}
The purely long-range ordered terms in the expression above define the free energy of the global polarization (order parameter):
\begin{equation}\label{Sfreeenergy}
F=\int{d{\bf x}}\Big(\frac{\alpha}{2}P^2+\frac{b}{4}P^4\Big),
\end{equation}
with the effective quadratic coefficient:
\begin{eqnarray}
  \alpha=a\!+\!b(2/d_p+1)\langle\delta{P}^2\rangle.\label{Salpha}
\end{eqnarray}
Here, $d_p$ is the dimension of the vector space of $\delta{\bf P}$. Minimizing the free energy yields  $P^2=-\alpha/b$ for $\alpha<0$ (ferroelectric phase) and $P^2=0$ for $\alpha>0$ (paraelectric phase). 

From the Lagrangian, the Euler-Lagrange equation of motion with respect to the polarization fluctuation $\delta{\bf P}$ is
\begin{equation}
\big[m_p\partial_t^2+\gamma\partial_t-g\nabla^2+a+b(2/d_p+1){P}^2+b\delta{P}^2\big]\delta{\bf P}={\bf E}_{\rm th}(t,{\bf R}).  \label{SPFEq}
\end{equation}
Here, we have introduced a damping rate $\gamma=0^+$ and a  thermal noise field ${\bf E}_{\rm th}(t,{\bf R})$ that  obeys the fluctuation-dissipation theorem:
\begin{equation}
\langle{\bf E}_{\rm th}(\omega,{\bf q}){\bf E}_{\rm th}(\omega',{\bf q}')\rangle=\frac{\gamma\hbar\omega(2\pi)^4\delta({\bf q}-{\bf q'})\delta(\omega-\omega')}{\tanh[\hbar\omega/(2k_BT)]},  
\end{equation}
with the four-momentum $p=(\omega,{\bf q})$ being the Fourier space of the space-time coordinate $x=(t,{\bf R})$. Under the mean-field approximation, combined with the assumption of isotropic fluctuations, yielding:
\begin{eqnarray}
 \delta P^2 \delta {\bf P} \rightarrow \Big(\frac{2}{d_p}+1\Big)\langle\delta P^2 \rangle \delta {\bf P},
\end{eqnarray} 
or 
\begin{eqnarray}
 \delta P^2 \delta {\bf P}&=&\sum_{i}\delta P^2 \delta {P}_i{\bf e}_i=\sum_{i}\delta P^2_i \delta {P}_i{\bf e}_i+\sum_{i{\ne}j}\delta P^2_j \delta {P}_i{\bf e}_i\rightarrow3\sum_{i}\langle\delta P^2_i\rangle \delta {P}_i{\bf e}_i+\sum_{i{\ne}j}\langle\delta P^2_j\rangle \delta {P}_i{\bf e}_i\nonumber\\
 &=&2\sum_{i}\langle\delta P^2_i\rangle \delta {P}_i{\bf e}_i+\sum_{i}\langle\delta P^2\rangle \delta {P}_i{\bf e}_i=\frac{2}{d_p}\sum_{i}\langle\delta P^2\rangle \delta {P}_i{\bf e}_i+\sum_{i}\langle\delta P^2\rangle \delta {P}_i{\bf e}_i=\Big(\frac{2}{d_p}+1\Big)\langle\delta P^2 \rangle \delta {\bf P},
\end{eqnarray}
the thermally averaged polarization fluctuation is given by:
\begin{equation}
  \langle{\delta{P}^2}\rangle=\!\!\int\!\!\frac{{\hbar}d{\bf q}}{2m_p(2\pi)^3}\frac{2n_B[\hbar\omega_{\bf q}(a,P^2,\langle\delta{P^2}\rangle)]+1}{\omega_{\bf q}(a,P^2,\langle\delta{P^2}\rangle)},\label{SPF}
\end{equation}  
where $\omega_{\bf q}$ is the energy dispersion of the polar mode:
\begin{equation}
\hbar \omega_{\bf q}(a,P^2,\langle\delta{P^2}\rangle)=\hbar \sqrt{\Delta^2(a,P^2,\langle\delta{P^2}\rangle)+{g q^2}/{m_p} },
\end{equation}
with excitation gap: 
\begin{eqnarray}\label{Sgap}
\Delta(a,P^2,\langle\delta{P^2}\rangle)\!=\!\big[a\!+\!b(2/d_p+1){P}^2+b(2/d_p+1)\langle\delta{P}^2\rangle\big]^{1/2}\!/\!\sqrt{m_p}.
\end{eqnarray}
While we have employed the fluctuation–dissipation theorem at this stage, the exactly same result can be rigorously derived using the fundamental path-integral formalism within the Matsubara (imaginary-time) representation~\cite{yang2024microscopic}, ensuring full consistency with the equilibrium quantum statistical mechanics.

Because the polar mode is bosonic, the fluctuation $\langle\delta{P}^2(T)\rangle\propto{n_B[\hbar\omega_{\bf q}(a,P^2,\langle\delta{P^2}\rangle)]}$ consists of the thermal fluctuations $\langle\delta{P}^2_{\rm th}(T)\rangle$ and zero-point fluctuations $\langle\delta{P}^2_{\rm zo}\rangle$. The thermal part vanishes at zero temperature and increases with temperature, while the zero-point part remains finite even at absolute zero. We note that the bare parameters such as $a$, which can in principle be obtained from the first-principles calculations,  are experimentally unobservable. Only after incorporating zero-point fluctuations, $a$ becomes  the experimentally accessible zero-temperature parameter $a_0$, and further applying the renormalization by thermal fluctuations leads to the free-energy parameter $\alpha(T)$ at finite temperatures.

At zero temperature, summarizing Eqs.~(\ref{SPF}) and~(\ref{Salpha}), the renormalization proceeds via:
\begin{eqnarray}\label{Sa0}
  a_0&=&a+b(2/d_p+1)\langle\delta{P}^2_{\rm zo}\rangle,\nonumber\\
  b_0&=&b,
\end{eqnarray}  
where the zero-point fluctuations:
\begin{equation}\label{Szo}
 \langle{\delta{P}^2_{\rm zo}}\rangle=\int\frac{\hbar}{2m_p}\frac{1}{\omega_{\bf q}(a,P^2_0,\langle\delta{P^2_{\rm zo}}\rangle)}\frac{d{\bf q}}{(2\pi)^3},  
\end{equation}
and the zero-temperature long-range ordered polarization: $P_0^2=-a_0/b_0$ for $a_0<0$ and $P_0^2=0$ for $a_0>0$. 

After the zero-point renormalization, the thermal contribution at finite $T$ gives:
\begin{eqnarray}
  \alpha(T)&=&a_0+b_0(2/d_p+1)\langle\delta{P}^2_{\rm th}(T)\rangle,\label{Saaaa}
\end{eqnarray}
where the thermal fluctuations are determined by substracting the zero-temperature part from Eq.~(\ref{SPF}) and are written as 
\begin{equation}
\langle{\delta{P}^2_{\rm th}(T)}\rangle=\int\frac{\hbar}{2m_p}\frac{2n_B[\hbar\omega_{\bf q}(a,P^2,\langle\delta{P^2}\rangle)]}{\omega_{\bf q}\big(a_0,P^2,\langle\delta{P_{\rm th}^2}(T)\rangle\big)}\frac{d{\bf q}}{(2\pi)^3},   \label{Sttf}
\end{equation}  
and finite-temperature long-range ordered polarization: $P^2=-\alpha(T)/b_0$ for $\alpha(T)<0$ and $P^2=0$ for $\alpha(T)>0$.

For accurate predictions for finite-temperature properties, if experimentally measured zero-temperature parameters are available, one may directly perform the thermal renormalization using Eq.~(\ref{Saaaa}). However, when only bare parameters from first-principles calculations are accessible, incorporating the zero-point renormalization is essential for quantitatively reliable predictions of the finite-temperature dielectric properties and critical behavior. In the main text, we directly use the experimentally measured
zero-temperature parameters~\cite{yang2024microscopic,rowley2014ferroelectric}.

The finite-temperature relative permittivity  is given by 
\begin{equation}
\varepsilon(T)=1+\frac{1/\varepsilon_0}{\alpha(T)+b_0P^2(1+2\cos^2\phi)}, 
\end{equation}
with $\phi$ being the angle between the polarization direction and the measurement field direction. Owing to the defect–polarization orientation coupling in the Hamiltonian, the polarization direction is thermally distributed rather than fixed, and the experimentally measured permittivity must be averaged over the thermal distribution of orientations, approximately yielding 
\begin{equation}
\varepsilon(T)=1+\frac{1/\varepsilon_0}{\alpha(T)+b_0P^2(T,x)[1\!+\!2\langle\cos^2(\theta-\phi)\rangle]}, 
\end{equation}
where the angular factor accounts for the thermal distribution of polarization orientations,
\begin{eqnarray}
1\!+\!2\langle\cos^2(\theta-\phi)\rangle&=&1\!+\!2\frac{\int\sin\theta{d}\theta\cos^2(\theta\!-\!\phi)e^{\beta\kappa\cos\theta}}{\int\sin\theta{d}\theta{e^{\beta\kappa\cos\theta}}}=2-W(T)+(3W(T)-1)\cos^2\phi,   
\end{eqnarray}
 with the orientational thermal distribution  $W(T)=\langle\cos^2\theta\rangle=1-\frac{2}{\beta\kappa}[\coth(\beta\kappa)-\frac{1}{\beta\kappa}]$.

In the present study, we set $\theta = 0.304\pi$, corresponding to the polarization oriented along the 
[100] direction and the measurement field along [111]. This choice reproduces well the measured data reported in Ref.~\cite{10.1103/physrevlett.39.1158}.

\subsection{Parameters in the simulation}

In our simulation, the model parameter $a_0(0)=22.5\times10^{-5}/\varepsilon_0$ is taken from the experimentally measured zero-temperature value for undoped KTaO$_3$~\cite{rowley2014ferroelectric,yang2024microscopic}. The polarization inertia is given by $m_p=\frac{\Omega_{\rm cell}}{\sum_iQ_i^2/M_i}$~\cite{sivasubramanian2004physical,tang2022excitations}, where $M_i$ and $Q_i$ denote the ionic masses and charges in the unit cell of volume $\Omega_{\rm cell}$, respectively. The gradient stiffness is taken as $\sqrt{{g}/{m_p}}=28.69~${\AA}/ps, as derived in Refs.~\cite{rowley2014ferroelectric,yang2024microscopic} based on the data from inelastic neutron~\cite{shirane1967temperature} and Raman scattering~\cite{fleury1968electric} measurements at 4~K. As derived in the main text, for small Nb doping, the doping-dependent parameter $a_0(x)$ scales as:
\begin{equation}
a_0(x)=a_0(0)\times(1-x/x_c),~~~~~~x_c=\delta{a}/a_0(0),
\end{equation}
and the experimentally determined quantum critical composition $x_c=0.008$. As noted in the main text, for each doping level, we determined $b_0$ and $\kappa$ by fitting to experimental data~\cite{10.1103/physrevlett.39.1158} in the low-temperature regime ($T<20~$K), where thermal effects of collective excitations are negligible. The fitted values are presented in Fig.~\ref{Sfig}.

Once these low-temperature parameters are determined, at each Nb doping level, we are able to quantitatively reproduce the experimentally measured dielectric properties over the entire temperature range (including the critical behavior near the phase transition), capturing the evolution from the zero-temperature condensed ground state, through the symmetry-broken ferroelectric phase, across the transition temperature, and into the paraelectric regime.

\begin{center}
\begin{figure}
    \includegraphics[width=12.7cm]{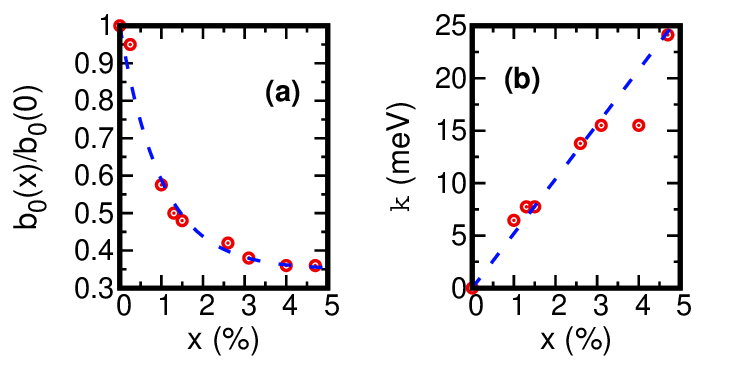}
    \caption{Fitted parameters  $b_0(x)$ (left panel) and $\kappa(x)$ (right panel) as functions of Nb composition $x$. Here,  $b_0(0)=0.37\times(2\pi)^3/\varepsilon_0$ (undoped case). The fits yield $b_0(x)/b_0(0)\approx0.65\exp(-100x)+0.35$, showing a rapid suppression of the anharmonic coefficient with Nb doping followed by saturation, consistent with a self-consistent-like renormalization of the polarization nonlinearity. 
    In contrast,  $\kappa(x)$ increases approximately linearly with $x$, reflecting the enhanced polarization–defect orientational coupling as the dopant concentration rises.}
    \label{Sfig}
\end{figure}
\end{center}

The schematic illustration in Fig.~1 of the main text is generated from a phenomenological minimal model introduced solely for qualitative and illustrative purposes, rather than for quantitative accuracy. The model is constructed from a Hamiltonian analogous in form to Eq.~(1). Specifically, we consider
\begin{align}
  &H=\int{d{\bf r}}\Big[\frac{1}{2}a_s|\phi({\bf r})|^2+\frac{1}{4}|\phi_{\rm sp}({\bf r})|^4+\frac{1}{2}(\nabla\phi({\bf r}))^2\Big].
\end{align}
The system is discretized on a two-dimensional Cartesian grid of $61\times61$ mesh points with periodic boundary conditions. To mimic local structural disorder, 300 defect-induced distortions are introduced at random lattice sites by fixing $\phi({\bf r}_i)=2$. The ground state is subsequently obtained by minimizing $H$. To emulate distinct dielectric regimes associated with different doping levels, we consider: (i) a classical paraelectric limit with a large positive quadratic coefficient $a_s \gg 0$; (ii) a quantum paraelectric limit with $a_s = 0^+$; and (iii) an intermediate regime at moderate $a_s$.

\section{Superconducting model}

In this section, we present the detailed superconducting model. As mentioned in the main text, the onset temperature of superconductivity is given within BCS theory by~\cite{bardeen1957microscopic,schrieffer1964theory} 
\begin{equation}\label{SCSC1}
T_{\rm os}^{\rm SC}=1.134{\bar \Omega}_{c}\exp(-1/\lambda),
\end{equation}
where $\lambda = -D\langle g_{\mathbf{kk'}}\rangle_F$ is the dimensionless coupling constant, $\langle\dots\rangle_F$ denotes the Fermi-surface average, $D$ is the electronic density of states for 2DEG formed at KTaO$_3$ interface, ${\Omega}_c$ is the cutoff energy of the attractive interaction, lying between the soft-phonon energy ${\bar \omega}_{q=0}$ and ${\bar \omega}_{q=2k_F}$ (see the following section), and we adopt the mean value ${\bar \omega}_{q=k_F}$.  Evaluating the pairing potential for soft-phonon exchange proposed in the main text yields a simplified form 
\begin{equation}\label{SCSC2}
\lambda=c_0\langle|{\bf k-k'}|^2/{\bar \omega}^{2}_{\bf k-k'}\rangle_F,
\end{equation}
where $c_0$ is a fixed constant. Experimentally, the soft-phonon gap of bulk KTaO$_3$ softens to $\Delta_{\rm sp}=3.79$ meV below 20 K~\cite{cheng2023terahertz,74d5-4hsw}, and the phonon velocity is $v=57~$\AA/ps~\cite{rowley2014ferroelectric}. Assuming approximate degeneracy of the three $t_{2g}$ orbitals at the KTaO$_3$(111) interface~\cite{10.1038/s41467-023-36309-2}, the Fermi wavevector is $k_F=\sqrt{2\pi n_{\rm 2D}/3}$, with $n_{\rm 2D}$ the interfacial carrier density. From the reported BCS coupling constant $\lambda=0.26$ at $n_{\rm 2D}=10^{14}$ cm$^{-2}$ for the EuO/KTaO$_3$(111) interface~\cite{10.1038/s41467-023-36309-2}, we estimate $c_0\approx490~$\AA$^2\cdot$meV$^2$. Then, the theoretically predicted superconducting $T_{\rm os}^{\rm SC}$ at KTaO$_3$ (111) interface as a function of carrier density is plotted in Fig.~3a in the main text, showing quantitative agreement with previous experimental measurements at EuO/KTaO$_3$(111) interfaces~\cite{10.1038/s41467-023-36309-2}. 

To quantitatively analyze the Nb doping effects, as mentioned in the main text, the renormalized soft-phonon gap ${\bar \Delta}_{\rm sp}(x)$ exhibits the same doping dependence as the polar-mode gap, such that in the low-temperature limit one has ${\bar \Delta}^2_{\rm sp}(x) = {\Delta}^2_{\rm sp}(x)$ for $x < x_c$ and ${\bar \Delta}^2_{\rm sp}(x) = -\frac{2}{3}{\Delta}^2_{\rm sp}(x)$ for $x > x_c$, i.e., ${\bar \Delta}_{\rm sp}(x)$ coincides with ${\Delta}_{\rm sp}(x)$ in the paraelectric phase but changes sign and magnitude in the ferroelectric phase due to the inversion of the soft-mode curvature. Here, ${\Delta}_{\rm sp}(x)={\Delta}_{\rm sp}(0)(1-x/x_c)$, as derived in the main text. Then, substituting the  ${\bar \Delta}_{\rm sp}(x)$ into Eqs.~(\ref{SCSC1}) and~(\ref{SCSC2}), the onset temperature of superconductivity $T_{\rm os}^{\rm SC}$ as a function of Nb doping, is plotted in Fig.~3(b) of the main text.  Our approach therefore demonstrates a direct link between ferroelectric quantum criticality and interfacial superconductivity at quantum paraelectric KTaO$_3$   interfaces. This connection arises from the role of soft phonons in mediating the pairing interaction, leading to an enhanced superconductivity as the system is tuned closer to the ferroelectric quantum critical point with the soft-phonon softening. 

Finally, it should be noted that in the present work we focus on the onset temperature $T_{\rm os}^{\rm SC}$ of superconductivity (i.e., gap-vanishing temperature), rather than the zero-resistance superconducting transition temperature $T_c^{\rm SC}$. The latter, in a two-dimensional system, requires a further inclusion of Berezinskii–Kosterlitz–Thouless (BKT) physics. However, as shown in our recent calculations~\cite{yang2025efficient}, in clean systems with small effective mass, the superfluid stiffness is large, and can  strongly suppress superconducting phase fluctuations. Consequently, the onset temperature and the BKT transition temperature become nearly coincident, allowing the onset temperature to serve as a reliable estimate of the true superconducting transition scale.

\subsection{Cutoff energy of the attractive interaction}

In this part, we discuss the cutoff energy of the attractive interaction. We start from the phonon-mediated pairing potential,
\begin{equation}\label{SSPS}
g_{\mathbf{k}\mathbf{k'}}(\Delta\xi,q)
=\frac{{\bar \omega}_{\mathbf{q}}|D^{\rm sp-e}_{\mathbf{q}}|^2}
{(\Delta\xi)^2-{\bar \omega}_{\mathbf{q}}^{2}},
\end{equation}
where $\Delta\xi=\xi_{\mathbf{k},n}-\xi_{\mathbf{k'},n'}$ and $\mathbf{q}=\mathbf{k}-\mathbf{k'}$. When $|\Delta \xi|<{\bar \omega}_{\mathbf{q}}$, the denominator of Eq.~(\ref{SSPS}) is negative and thus $g_{\mathbf{kk'}}<0$, indicating an attractive interaction. In contrast, for $|\Delta \xi|>{\bar \omega}_{\mathbf{q}}$, the interaction becomes repulsive or ineffective. Therefore, for a given momentum transfer $\mathbf{q}$, pairing is attractive only within the energy window $|\Delta \xi|\lesssim {\bar \omega}_{\mathbf{q}}$. Since the soft-phonon dispersion ${\bar \omega}_{\mathbf{q}}=\sqrt{{\bar \Delta}_{\rm sp}^2+v^2q^2}$ increases monotonically with $q$, and the momentum transfer satisfies $0<q\le2k_F$ for states near the Fermi surface, the relevant phonon frequencies span the range between ${\bar \omega}_{q=0}={\bar \Delta}_{\rm sp}$ and ${\bar \omega}_{q=2k_F}=\sqrt{{\bar \Delta}_{\rm sp}^2+v^2(2k_F)^2}$.

The onset temperature $T^{\rm SC}_{\rm os}$ of superconductivity is determined by the linearized BCS gap equation: 
\begin{equation}
\Delta(\mathbf{k})=-\sum_{\mathbf{k}'} g_{\mathbf{kk'}}
\frac{\tanh\big(\xi_{\mathbf{k}'}/2T^{\rm SC}_{\rm os}\big)}{2\xi_{\mathbf{k}'}}\Delta(\mathbf{k}'),
\label{eq:linBCS}
\end{equation}
where $\xi_{\mathbf{k}}=\varepsilon_{\mathbf{k}}-\mu$. According to the Cooper instability, the summation is restricted to scattering processes with $g_{\mathbf{kk'}}<0$, corresponding to momentum transfers that yield an attractive interaction. 

Separating the sum over $\mathbf{k}'$ into an energy integral and an angular integral,
\begin{equation}
\sum_{\mathbf{k}'}\;\to\; D\int d\xi'\int\frac{d{\bar \omega}_{\hat{\mathbf{k}}'}}{S_d},
\qquad
\mathbf{q}=\mathbf{k}-\mathbf{k}' ,\;\; 0\le q\le 2k_F,
\end{equation}
with $D$ the density of states at the Fermi level and $S_d$ the solid angle in $d$ dimensions. Taking $\Delta(\mathbf{k})$ as a constant on the Fermi surface ($s$–wave), the angular dependence and slowly varying parts of $g_{\bf kk'}$ are absorbed into a non-negative weight function $W(q)$, and  Eq.~(\ref{eq:linBCS}) becomes
\begin{equation}
1\simeq-D\int_{0}^{2k_F}{dq}W(q)g(q)\,\,
\int_{0}^{{\bar \omega}_q}\!\frac{d\xi}{\xi}\,\tanh\!\Big(\frac{\xi}{2T^{\rm SC}_{\rm os}}\Big).
\label{eq:preT}
\end{equation}
Near $T=T^{\rm SC}_{\rm os}$, the standard approximation,
\begin{equation}
\int_{0}^{{\bar \omega}_q}\frac{d\xi}{\xi}\tanh\!\Big(\frac{\xi}{2T^{\rm SC}_{\rm os}}\Big)
\simeq \ln\!\Big(\frac{1.134\,{\bar \omega}_q}{T^{\rm SC}_{\rm os}}\Big)
\end{equation}
applies, where the numerical factor $1.134=(2e^\gamma/\pi)$ comes from the Matsubara summation ($\gamma$ is the Euler constant). Substituting this into Eq.~(\ref{eq:preT}) yields
\begin{equation}
1 \simeq -D\int_{0}^{2k_F}\!\!dq\,W(q)\,g(q)\,
\ln\!\Big(\frac{1.134\,{\bar \omega}_q}{T^{\rm SC}_{\rm os}}\Big).
\label{eq:logavg}
\end{equation}
Defining the dimensionless coupling constant and logarithmic cutoff as
\begin{equation}
\lambda \equiv -D\,\frac{\displaystyle\int_{0}^{2k_F}\!dq\,W(q)\,g(q)}{\displaystyle\int_{0}^{2k_F}\!dq\,W(q)}\approx-D\langle g_{\mathbf{kk'}}\rangle_F ,
\qquad
\ln{\Omega}_c \equiv \frac{\displaystyle\int_{0}^{2k_F}\!dq\,W(q)\,\ln {\bar \omega}_q}{\displaystyle\int_{0}^{2k_F}\!dq\,W(q)} ,
\end{equation}
Eq.~(\ref{eq:logavg}) takes the standard BCS form
\begin{equation}
T^{\rm SC}_{\rm os} = 1.134\,{\Omega}_c\,e^{-1/\lambda}.
\end{equation}
Clearly, because ${\bar \omega}_q$ increases monotonically with $q$ between ${\bar \omega}_{q=0}={\bar \Delta}_{\rm sp}$ and ${\bar \omega}_{q=2k_F}=\sqrt{{\bar \Delta}_{\rm sp}^2+v^2(2k_F)^2}$, the logarithmic average necessarily satisfies
\begin{equation}
{\bar \omega}_{q=0}\ \le\ {\Omega}_c\ \le\ {\bar \omega}_{q=2k_F}.
\end{equation}
Following the treatment of the coupling constant by taking average of pairing potential on the fermi surface, after similarly evaluating $\ln{\Omega}_c=\langle\ln {\bar \omega}_{\bf k-k'}\rangle$, we find that ${\Omega}_c$ can be well approximated by the intermediate value ${\bar \omega}_{k_F}$.

\end{appendix}
\end{widetext}

%
\end{document}